# Dye-Sensitized Ternary Copper Chalcogenide Nanocrystals: Optoelectronic Properties, Air Stability and Photosensitivity

*Sonam Maiti[1,2], Santanu Maiti[2], Ali Hossain Khan[3,4], Andreas Wolf[5], Dirk Dorfs[5], Iwan Moreels[3,4],*

*Frank Schreiber[2,6] and Marcus Scheele[1,6] ∗*

[1] Institute of Physical and Theoretical Chemistry, University of Tübingen, Auf der Morgenstelle 18, 72076 Tübingen, Germany.

[2] Institute of Applied Physics, University of Tübingen, Auf der Morgenstelle 10, 72076 Tübingen, Germany

[3] Istituto Italiano di Tecnologia, Via Morego 30, 16163 Genova, Italy

[4] Department of Chemistry, Ghent University, Krijgslaan 281-S3, 9000 Gent, Belgium

[5] Institute of Physical Chemistry and Electrochemistry, Leibniz Universität Hannover, Callinstr. 3A, 30167 Hannover, Germany

[6] Center for Light-Matter Interaction, Sensors & Analytics LISA+, University of Tübingen, Auf der Morgenstelle 15, 72076 Tübingen, Germany.

## Abstract

We report on the effect of ligand exchange of $Cu_2Se_yS_{1-y}$ as well as $Cu_2Se$ nanocrystals (NCs) with the organic π-system Cobalt β-tetraaminophthalocyanine (CoTAPc) and analyse changes in the structural, optical as well as electric properties of thin films of these hybrid materials. A strong ligand interaction with the surface of the NCs is revealed by UV/vis absorption and Raman spectroscopy. Grazing-incidence small-angle X-ray scattering studies show a significant contraction in the interparticle distance upon ligand exchange. For copper-deficient $Cu_{2-x}Se$, this contraction has a negligible effect on electric transport, while for copper-deficient $Cu_{2-x}Se_yS_{1-y}$, the conductivity increases by eight orders of magnitude and results in metal-like temperature-dependent transport. We discuss these differences in the light of varying contributions of electronic vs. ionic transport in the two materials and highlight their effect on the stability of the transport properties under ambient conditions. With photocurrent measurements, we demonstrate high optical responsivities of 200-400 A/W for CoTAPc-capped $Cu_2Se_yS_{1-y}$ and emphasize the beneficial role of the organic π-system in this respect, which acts as an electronic linker and an optical sensitizer at the same time.

# Introduction

Copper chalcogenide nanocrystals (NCs) have become a subject of intense research as possible alternatives to the more toxic Cd or Pb-based counterparts for optoelectronic applications such as solar cells or photocatalysts, but also as thermoelectric converters, gas sensors, optical filters, superionic conductors and electro-optical devices.[1-13] A majority of these investigations focused on the binary compounds copper selenide ($Cu_2Se$) and copper sulphide ($Cu_2S$) with variable Cu(I) deficiency (e.g., $Cu_{2-x}Se$, $0.00 \leq x \leq 0.6$) and tailored charge carrier concentration as well as charge carrier concentration dependent localized plasmon resonance frequency (LSPR).[11, 14-23] Introducing Cu(I) vacancies is readily afforded by oxidizing parts of the chalcogenide sublattice into the (-I)-state, which leads to a loss of Cu(I)-ions and the release of free holes in the NC core.[24] More recently, the ternary alloy, $Cu_2Se_yS_{1-y}$, has been studied to some extent, for instance as precursor in the synthesis of $Cu_2ZnSn(Se_yS_{1-y})_4$ NCs with relevance for photovoltaic applications.[7, 8, 25-30] Particular, with respect to the oxidation-sensitive $Cu_2Se$ system, it has been argued that $Cu_2Se_yS_{1-y}$ may have similar optoelectronic properties, however with improved stability in air.[31, 32] For the binary, copper-deficient copper chalcogenide NCs, $Cu_{2-x}Se$ and $Cu_{2-x}S$, several studies have been conducted to increase electronic coupling in thin films of these materials, for instance by removing the native ligand with smaller molecules or anions, by thermal decomposition of the insulating ligand sphere or by thermal doping.[3, 5, 19, 20, 33-40] Specifically for $Cu_{2-x}Se$ NCs, several groups have reported that high electric conductivities (up to 25 $Scm^{-1}$) may also be obtained without such post-synthetic ligand exchange, indicating that charge carrier transport in these NCs potentially follows a different mechanism than in copper sulphide where ligand exchange is usually necessary.[41-44] From the perspective of tailoring the optoelectronic properties of copper chalcogenide NCs by their surface chemistry, such ligand-independent transport characteristics are undesirable. However, the electric conductivities in $Cu_2Se$ NCs are often found to be larger than in comparable $Cu_2S$ NCs.[37, 45]

The present study is motivated by the hypothesis that alloying $Cu_2S$ into $Cu_2Se$ may combine the ligand-tunable optoelectronic of pure $Cu_2S$ with the high electric conductivity of pure $Cu_2Se$. To this end, we compare the optical and electrical properties of $Cu_2Se$ and $Cu_{2-x}Se$ with $Cu_2Se_yS_{1-y}$ and $Cu_{2-x}Se_yS_{1-y}$, assess the different sensitivity to oxidation in air and the effect of ligand exchange with the organic π-system Cobalt β-tetraaminophthalocyanine (CoTAPc). We show that only the ternary alloy exhibits stable electric transport properties in air. Electrical conductivities > 1 Scm$^{-1}$ and an increasing resistivity with increasing temperature indicate highly efficient charge carrier transport in CoTAPc-functionalized $Cu_{2-x}Se_yS_{1-y}$ NC thin films. We demonstrate an optical responsivity of 400 A/W under 637 nm photoexcitation which is an exceptionally large photosensitivity for this material. We argue that this is enabled by the hybrid nature of the presented material, in which the organic π-system acts as the photosensitizer and the network of NCs provides the channel for fast transport of the photoexcited charges.

## 2. Methods

### 2.1. Synthesis of Cu$_2$Se nanocrystals:

A synthesis method adapted from Deka et al. has been used to produce quasi-spherical Cu$_2$Se NCs.[43] Standard Schlenk line techniques were applied during synthesis and purification. A mixture of 15 mL 1-Octadecene (ODE) and 15 mL Oleylamine (OLm) is degassed under vacuum -at 115 °C for 3 h. The mixture is cooled to room temperature, and under argon flow 297 mg Copper(I) chloride (CuCl, 3 mmol) is added. For an additional 15 min, the mixture is heated to reflux under vacuum. Subsequently, the flask is filled with argon and the temperature is raised to 300 °C in 5-6 min. The selenium (Se) precursor solution is prepared by dissolving Se (117 mg, 1.5 mmol) in degassed OLm (9 mL) and refluxing under vacuum for 30 min (115 °C). The flask is again filled with argon and the mixture is left stirring at 190-200 °C until all Se is dissolved. Upon dissolution, the temperature is raised to 230 °C for 20 min. To transfer the precursor solution with a glass syringe, the solution is cooled to 150 °C. The Se solution is rapidly injected into the copper precursor solution. The temperature of the reaction mixture is allowed to recover to 290 °C within approximately 2-3 min. The reaction is quenched 15 min after the injection. At 150 °C, toluene (20 mL) is injected to prevent agglomeration. The particles are precipitated from the growth solution with ethanol (20 mL) and methanol (10 mL) and centrifugation (3700g, 20 min). The precipitate is resuspended in toluene (20 mL) by ultrasonication (5 min). After 12 h the solution is centrifuged again (3700g, 20 min) to remove aggregates. The supernatant is collected and used for all further experiments.

**2.2. Synthesis of $Cu_2Se_yS_{1-y}$ nanocrystals:**

The Se-precursor solution is prepared according to the reported method of Lesnyak et al.[2] 158 mg of Se powder (2 mmol) is mixed with 1 mL of 1-dodecanethiol (DDT) and 1 mL of OLm and is heated for 1 h at 100 °C under nitrogen atmosphere. The resultant, brown alkyl ammonium selenide solution is cooled to room temperature and stored in a nitrogen filled glass vial.

In a three-neck round-bottom flask, 262 mg of Copper(II) acetylacetonate [$Cu(acac)_2$] (1 mmol) is mixed with 3.5 mL of DDT and 10 mL of OLm and the mixture is degassed under vacuum with vigorous stirring at 70 °C for 1 h. Next, the flask is filled with nitrogen and quickly heated to 220 °C. After complete dissolution, $Cu(acac)_2$ forms a clear yellow solution. At this temperature, a mixture of 0.5 mL of the Se-precursor with 1.5 mL of DDT is swiftly injected into the flask leading to immediate color change from yellow to brown. The reaction mixture is kept at 220 °C for 30 min. The nanocrystals are precipitated under inert gas atmosphere by centrifugation of the crude reaction mixture with subsequent dissolving of the precipitate in chloroform.

**2.3. Thin-Film Preparation and Ligand Exchange:** NC thin films were prepared by assembly at the dimethylsulfoxide/$N_2$ interface under inert conditions in a glovebox. The fabrication process and ligand exchange were carried out in a home-built Teflon chamber according to our previously reported method.[5]

**2.4. Instrumentation:** Scanning transmission electron microscopy (STEM, Hitachi SU 8030 microscope operating at 30 kV) is employed to determine the particle size and shape. Optical measurements are performed on solid state films on glass substrates using an UV–vis–NIR spectrometer (Agilent Technologies, Cary 5000). Grazing-incidence small-angle X-ray scattering (GISAXS)[6, 46-49] is carried out with a laboratory instrument (Xeuss 2.0, Xenocs, France) using Cu $K_\alpha$ radiation ($\lambda$ =1.5418 Å). The samples are probed with a focused X-ray beam of size 0.5 x 0.5 $mm^2$ at an incidence angle of $0.22^0$. The GISAXS images are collected with a 2D Pilatus 300K, having 487 x 619 pixels. The detector is placed at a distance of 2496 mm, determined using Ag-behenate as reference sample. X-ray diffraction (XRD) data from the sample is collected in a laboratory source (Cu $K_\alpha$; GE Inspection Technologies, Germany). Raman spectra are acquired using a Horiba Jobin Yvon Labram HR 800 spectrometer with a CCD-1024 × 256-OPEN-3S9 detector. Excitation for Raman is performed using a He:Ne laser (wavelength 633 nm). Electrical measurements are performed in a glovebox at room temperature with a homemade probe station using a Keithley 2634B dual source-meter unit, controlled by the included test script builder program. The NC films after ligand exchange are deposited on commercially available bottom-gate, bottom-contact transistor substrates (Fraunhofer Institute for Photonic Microsystems, Dresden, Germany) with interdigitated Au electrodes of 10 mm width and 2.5 µm channel length followed by annealing at 250 °C for 2 h under nitrogen atmosphere. The temperature-dependent charge transport properties as well as the photoresponse of the NC thin-films are measured in a Lake-Shore CRX-6.5K probe station at a pressure of $5\times10^{-6}$ mbar, equipped with a Keithley 2636B dual source-meter unit and a Lake Shore temperature controller (model 336). As an excitation source, single mode fiber-pigtailed laser diodes operated by a compact laser diode controller CLD1010 by Thorlabs were used: A 638 nm laser diode with a maximal output power of 70 mW and a 408 nm diode with a maximal output power of 30 mW. Losses to this theoretical optical power output due to scattering, inefficient coupling into the optical fiber, decollimation of the beam etc. were determined by a calibration sample and an optical power meter to obtain the total absorbed optical power at the sample surface.

**Results and Discussion**

The TEM images in **Figure 1a** and **1b** depict the morphologies of the as-synthesized alloyed $Cu_2Se_yS_{1-y}$ and $Cu_{2-x}Se$ NCs with relatively uniform size of 7.0 ± 0.8 nm and 12.2 ± 1.9 nm, respectively. All NCs appear well separated by the native OLm capping ligand. Powder X-ray diffraction reveals that the $Cu_2Se_yS_{1-y}$ are in the hexagonal phase (**Figure S1**).[8] We determine the composition of the alloyed NCs by energy-dispersive X-ray spectroscopy (EDX) as $Cu_{2.2}Se_{0.68}S_{0.32}$ (**Figure 1c**). We note that colloidal NCs often exhibit an excess of cations on the surface, such that the actual stoichiometry of the inner core may contain less copper than suggested by our EDX result. **Figure 1d** shows Raman spectra for solid state films of both NC materials coated with the native OLm ligand. In good agreement with previous reports, the most intense resonance peak is observed at around 260 cm$^{-1}$ for both samples, corresponding to the Se–Se stretch vibration.[21, 50] For the films composed of the ternary $Cu_2Se_yS_{1-y}$ NCs, two additional peaks are observed, one at 450 cm$^{-1}$ corresponding to the S-S stretching mode, and another peak at 368 cm$^{-1}$ due to the S-Se stretching mode.[51] Consistent with the EDX data, the intensity of the Se-Se band is much stronger than that of the S-S and S-Se stretching vibration, suggesting that the alloy is rich in selenium.

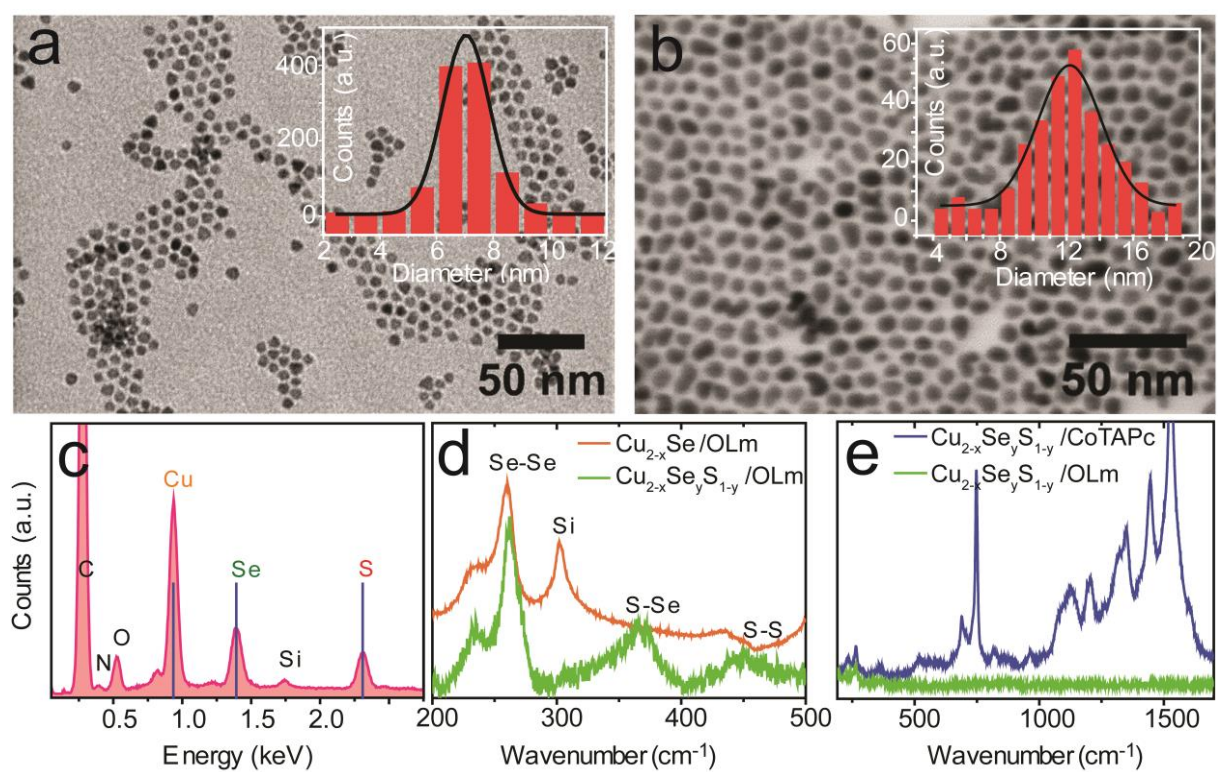

**Figure 1.** TEM images of **(a)** 7.0 ± 0.8 nm $Cu_2S_ySe_{1-y}$ NCs and **(b)** 12.2 ± 1.9 nm $Cu_2Se$ NCs. **(c)** EDX spectrum of the $Cu_2S_ySe_{1-y}$ NCs. **(d)** Raman spectra of as-prepared $Cu_2Se$ and $Cu_2Se_yS_{1-y}$ nanocrystal thin films (orange curve and green curve, respectively). **(e)** Raman spectra of CuSeS nanocrystal thin films before and after ligand exchange (green curve and blue curve, respectively).

To enhance chemical and electronic coupling in solid-state films of both NC materials, we exchange the native OLm ligand with the multidendate cross-linker CoTAPc. We choose this linker because earlier reports have shown that tetraaminophthalocyanines are suitable to replace oleylamine from the surface of $Cu_{1.1}S$ NCs and drastically improve charge carrier transport.[5] The ligand-exchanged NC films exhibit a smooth surface with an average height difference of 3-4 NCs and 1 NC monolayer for $Cu_2Se_yS_{1-y}$ and $Cu_2Se$, respectively (**Figure S2**). We monitor the effect of this ligand exchange by Raman spectroscopy in **Figure 1e**. (This Figure exemplifies the exchange for $Cu_2Se_yS_{1-y}$ NCs, but the same spectral features between 700 – 1700 $cm^{-1}$ are also obtained with $Cu_2Se$ after ligand exchange.) In accordance with previous studies, we interpret the strong bands appearing at 747, 1124, 1202, 1337, 1447, 1530 and 1605 $cm^{-1}$ with vibrational modes of CoTAPc, which is supporting evidence for the presence of the new linker in the NC film.[52] The peaks at 300 and 513 $cm^{-1}$ belong to the Si substrate. Fourier-transform infrared spectroscopy furthermore reveals significant changes after exposure to CoTAPc, most notably the disappearance of characteristic OLm vibrations at 1660 $cm^{-1}$ and 3350 $cm^{-1}$. (For details, see **Figure S3** in the Supporting Information.)

We determine the structural details of the ensemble of the two NC samples as well as the effect of ligand exchange with CoTAPc by GISAXS in **Figure 2**. The intense in-plane scattering truncation rods, extended along the $q_z$-direction indicate the formation of superlattices with long-range in-plane order. For OLm-capped $Cu_{2-x}Se$ NCs (**Figure 2a+c**), we find the first order in-plane correlation peak at $q_y$ = 0.043 $Å^{-1}$, a second order peak at $q_y$ = 0.078 $Å^{-1}$ and a barely visible third signal at $q_y ≈$ 0.9 $Å^{-1}$. These relative positions in $q_y$ can be interpreted as the formation of a hexagonal lattice (e.g. $q_1:q_2:q_3 = 1:\sqrt{3}:2$) with in-plane lattice constant a = 16.8± 0.1 nm. After ligand exchange with CoTAPc (**Figure 2b+c**), the in-plane correlation peaks shift to higher values, that is, smaller lattice constants, and we find $q_y$ = 0.048 $Å^{-1}$, 0.084 $Å^{-1}$ and a shoulder at 0.096 $Å^{-1}$. These values are again in agreement with a hexagonal lattice, but with a contracted in-plane lattice constant a = 15.1± 0.1 nm. We attribute the contraction of 1.7 nm to the replacement of OLm by the smaller CoTAPc ligand. The improved signal-to-background ratio after ligand exchange indicates a higher degree of long-range order as a result of cross-linking with the rigid organic π-system. With respect to the average particle diameter of 12.2± 1.9 nm (Figure 1b), the interparticle spacing before ligand exchange is 4.6±1.9 nm, which can be interpreted with two adjacent, non-intercalated ligand spheres of OLm. After cross-linking with CoTAPc, the interparticle spacing is reduced to 2.9±1.9 nm, which is equivalent to 1-2 times the molecular length of CoTAPc.

Since the exact binding mode of CoTAPc to the surface of the NCs is not known, the latter finding could either be explained with a side-on binding of stacks of CoTAPc or with head-to-tail cross-linking of a CoTAPc monolayer.

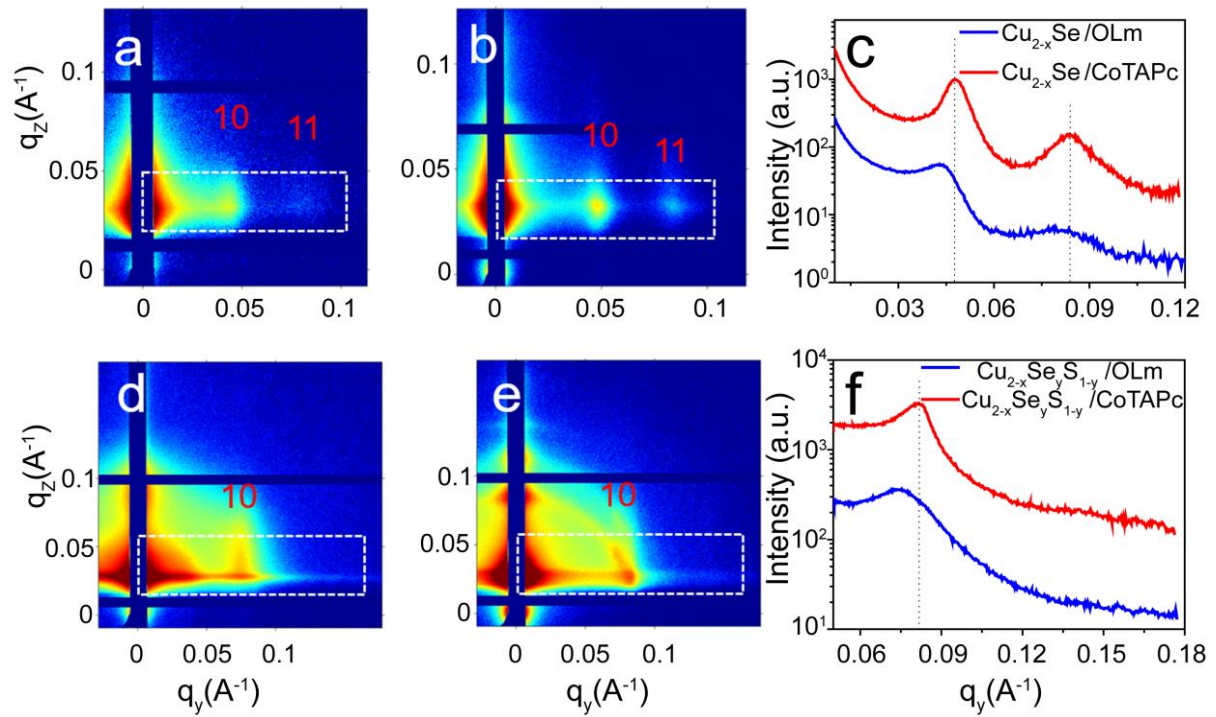

**Figure 2.** GISAXS patterns of self-assembled **(a)** $Cu_2Se$/OLm, **(b)** $Cu_2Se$/CoTAPc, **(d)** $Cu_2Se_yS_{1-y}$/OLm and **(e)** $Cu_2Se_yS_{1-y}$/CoTAPc films. (c) and (f) extracted line profiles from the corresponding GISAXS images on the left in (a)/(b) and (d)/(e), respectively, as a function of the in-plane scattering vector $q_y$. To improve the statistics of the line profiles, the ROI (white dotted box) was integrated along the $q_z$ direction.

Similar GISAXS patterns are also obtained for $Cu_{2-x}Se_yS_{1-y}$ before and after ligand exchange with CoTAPc (**Figure 2d+e**). The first-order correlation peak shifts from 0.074 Å$^{-1}$ with OLm functionalization to 0.082 Å$^{-1}$ after ligand exchange (**Figure 2f**). Under the assumption of a hexagonal lattice, this corresponds to a center-to-center distance between the NCs of 9.8±0.1 nm for OLm functionalization and 8.8 ± 0.1 nm for CoTAPc as the ligand. With a particle diameter of 7.0±0.8 nm, the interparticle distances are 2.8±0.8 nm for OLm and 1.8±0.8 nm for CoTAPc. The latter result may be viewed as indirect supporting evidence that CoTAPc binds preferentially in a head-to-tail cross-linking manner between the surfaces of two adjacent NCs as the width of the ligand sphere is precisely one molecular length of CoTAPc.

For charge transport studies, we deposit both copper chalcogenide NC films on silicon oxide substrates with pre-patterned Au contacts and record the two-point probe current–voltage (I–V) characteristics at room temperature. In **Figure 3a+b**, we focus on a comparison of the I/V characteristics of both materials before and after ligand exchange with CoTAPc and before/after oxidation by exposure to air. The left panels in **Figure 3** represent the characteristics of the ternary copper chalcogenide NCs, while the right panel characterizes the binary NCs. The color code is the same for both materials: green = OLm capping, reduced; yellow = OLm capping, oxidized; blue = CoTAPc capping, reduced; red = CoTAPc capping, oxidized. Oxidation leads to copper vacancies and a non-stoichiometric composition in copper selenide NCs and drastically increases the density of free holes, which manifests in degenerate p-type doping as well as the occurrence of an LSPR in the near-infrared (NIR).[10, 16, 22] Therefore, we monitor the degree of vacancy doping for both samples with vis/NIR absorption spectroscopy in **Figure 3c+d.** We note a broad band centred at 1250 nm for oxidized $Cu_{2-x}Se_yS_{1-y}$ (for both Olm and CoTAPc ligands), 1300 nm for oxidized OLm-capped $Cu_{2-x}Se$ and 1600 nm for oxidized $Cu_{2-x}Se$ capped with CoTAPc, which we interpret as LSP resonances. In the reduced state, both materials show a negligible LSPR signal below 2000 nm, indicative of a low carrier density and a near stoichiometric copper content. In both ligand exchanged samples, the HOMO-LUMO transition of CoTAPc invokes a strong absorption band between 600-800 nm. (See Supporting Information for the absorption spectrum of pure CoTAPc. **Figure S4**) Before ligand exchange, charge transport is poor in both materials (green curve) with conductivities on the order of $10^{-8}$ S cm$^{-1}$ for $Cu_2Se_yS_{1-y}$ and $10^{-4}$ S cm$^{-1}$ for $Cu_2Se$. After surface functionalization with CoTAPc, both materials behave similar and the conductivities increase dramatically to 1 S cm$^{-1}$ and 5 S cm$^{-1}$, respectively (blue curve). In contrast, when studying the effect of oxidation in air for several hours ($Cu_{2-x}Se$) or days ($Cu_{2-x}Se_yS_{1-y}$), we observe a different behavior for $Cu_{2-x}Se_yS_{1-y}$ *vs.* $Cu_{2-x}Se$. While the increase of copper vacancies has a negligible effect on the transport properties of OLm-capped $Cu_{2-x}Se_yS_{1-y}$, it increases the conductivity of the OLm-capped $Cu_{2-x}Se$ to 6 S cm$^{-1}$. Oxidizing the CoTAPc-capped NC films has no significant effect on the conductivity of either of the two samples (**Figure S5**).

The structural characterization in **Figure 2** demonstrates that the highly conductive OLm-capped Cu$_{2-x}$Se NCs are well-separated from each other (4.6±1.9 nm), such that necking and the formation of percolative pathways are an unlikely explanation for such efficient charge carrier transport. Similar widths of the OLm ligand sphere are also observed for Cu$_{2-x}$Se$_y$S$_{1-y}$ (2.8±0.8 nm, this work) and Cu$_{1.1}$S NCs (4.1±1.7 nm), which exhibit negligible conductivities (10$^{-8}$ S cm$^{-1}$ and 10$^{-9}$ S cm$^{-1}$, respectively).[5] We note that previous reports on drop-casted, oxidized OLm-capped Cu$_{2-x}$Se NCs revealed similar or even larger conductivities, corroborating our finding here that Cu$_{2-x}$Se NCs show uniquely different transport properties compared to Cu$_{2-x}$Se$_y$S$_{1-y}$ or Cu$_{1.1}$S NCs.[41-43] It is furthermore surprising that the significant contraction of the Cu$_{2-x}$Se NC ensemble by 1.7 nm upon ligand exchange has no effect on the conductivity (**Figure 3b** yellow *vs.* red curve). This speaks against electronic (hopping) conduction as the dominant transport mechanism in OLm-capped Cu$_{2-x}$Se NC thin films, which is strongly affected by a change of the hopping distance.[53] We hypothesize that ionic conduction of mobile copper ions may play a key role here. Very large ionic mobilities with diffusion constants > 10$^{-5}$ cm$^2$s$^{-1}$ and superionicity have been reported for Cu$_{2-x}$Se, which can result in electric conductivities > 1 Scm$^{-1}$.[10, 54] Although it is not immediately obvious how such ionic conductivity would be affected by the OLm ligand sphere, superionicity is an important feature of Cu$_{2-x}$Se NCs and likely to be responsible for the unusually large electric conductivities. This would explain why only a reduction in copper vacancies can significantly reduce the conductivity in OLm-capped Cu$_2$Se NCs. After surface-functionalization with CoTAPc, electronic conduction appears greatly improved, such that the conductivity is now only weakly affected by the density of copper vacancies (**Figure 3b** blue *vs.* red curve).

An alternative explanation which we briefly consider here involves the formation of conductive, percolative pathways consisting of copper oxide nanostructures. Oxidation in air of Cu$_2$Se NCs results in the release of Cu(I)-ions, which react with oxygen to copper oxide NCs.[16] The conductivity of some copper oxide phases (which are mostly semiconducting) is rather high, and it is possible that successive release of Cu(I) ions from the Cu$_2$Se NCs leads to a continuous network of this conductor. Once formed, charge transport across this network is expected to be unaffected by the addition of [Cu(CH$_3$CN)$_4$]PF$_6$, which is a powerful reducing agent for Cu$_{2-x}$Se NCs (via filling of Cu(I) vacancies), but not for copper oxide. However, we find that films of Cu$_{2-x}$Se/OLm NCs show a drastically reduced electric conductivity upon treatment with [Cu(CH$_3$CN)$_4$]PF$_6$, which speaks against this alternative explanation (see **Figure S6**).

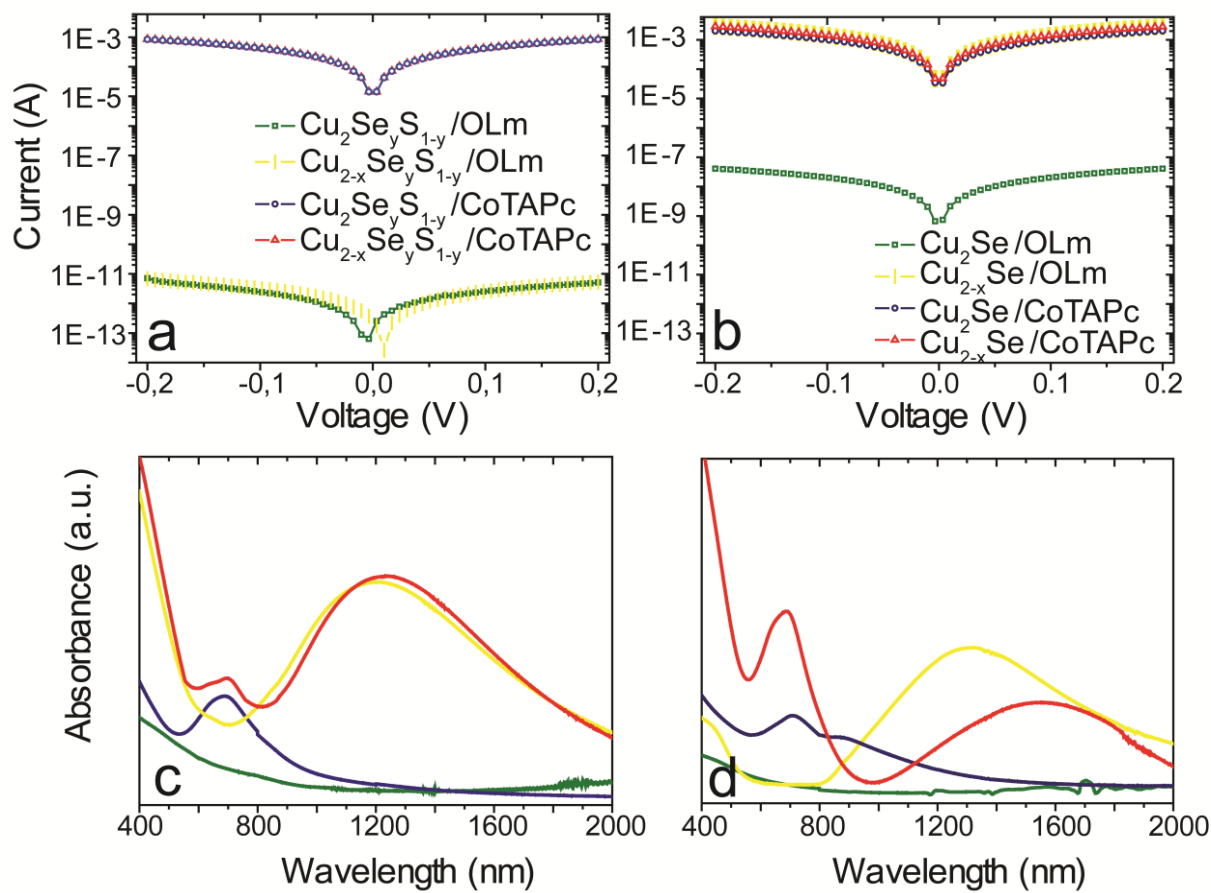

**Figure 3.** (a) Current-voltage (*I-V*) curves of Cu$_2$Se$_y$S$_{1-y}$ as well as Cu$_{2-x}$Se$_y$S$_{1-y}$ and (b) of Cu$_2$Se and Cu$_{2-x}$Se. (c) Corresponding optical absorption spectra of the ternary and (d) binary copper chalcogenides. The color code is the same in all four panels: green = Olm capping, reduced; yellow = Olm capping, oxidized; blue = CoTAPc capping, reduced; red = CoTAPc capping, oxidized.

The transport characteristics of the ternary Cu$_2$Se$_y$S$_{1-y}$ NC ensemble appears to be dominated by electronic conduction with a strong dependence on the hopping distance (**Figure 3a** yellow *vs.* red curve and green *vs.* blue curve) and weak dependence on the density of copper vacancies (**Figure 3a** green *vs.* yellow curve and blue *vs.* red curve). In view of a recent report on electric transport in similar Cu$_{1.1}$S

NC ensembles, the $Cu_2Se_yS_{1-y}$ NCs investigated here resemble much more that of the binary sulphides than the selenides.[5]

To further understand the electronic properties of CoTAPc-capped $Cu_2Se_yS_{1-y}$ NCs, we perform temperature-dependent resistivity measurements under high vacuum (**Figure 4a**). Throughout the entire temperature regime of 20-300 K, we observe monotonically increasing resistance with temperature, reminiscent of metallic behavior. Such characteristic is rarely observed in copper chalcogenide NC ensembles, and only in cases where electronic coupling is large enough to overcome the temperature-activated hopping regime.[34, 55]

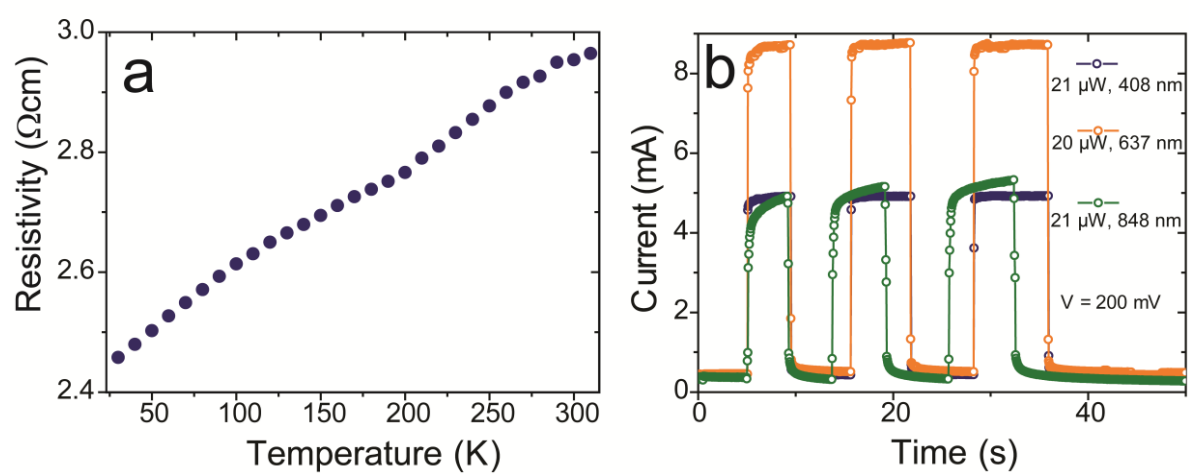

**Figure 4.** (**a**) Temperature-dependent resistivity of $Cu_2Se_yS_{1-y}$ functionalized with CoTAPc. (**b**) Time-dependent current at 200 mV of a $Cu_2Se_yS_{1-y}$/CoTAPc film during three excitation periods to 21 µW of 408 nm light (blue), 20 µW of 637 nm light (orange) and 21 µW of 848 nm (green).

In light of the tunable optical absorption properties, copper chalcogenide NCs are often considered for applications where light-to-electric current conversion is important, such as photovoltaics or photocatalysis. However, the photocurrent behaviour of these materials showed only moderate photosensitivities so far.[56-59] In **Figure 4b**, we display the ON/OFF photocurrent characteristics of $Cu_2Se_yS_{1-y}$/CoTAPc NCs during three excitation periods with 408 nm, 637 nm and 848 nm laser diodes at roughly the same direct optical power of ~20 µW. At a bias of 200 mV, we find a photocurrent of ~4 mA, ~ 8mA and 4 mA, respectively, corresponding to a responsivity of 200 A/W at 408 nm and 848 nm as well as 400 A/W at 637 nm. The reversibility of the transport characteristics after each excitation period indicates that the increased current is indeed a photocurrent and not, as recently observed for copper sulphide NCs, an irreversible photo-doping effect.[38] We explain such unprecedented optical responsivity with the presence of CoTAPc, which shows strong absorbance at all three excitation wavelengths (See Supporting Information). We suggest that CoTAPc sensitizes $Cu_2Se_yS_{1-y}$ NCs for the absorption of photons at these wavelengths to form singlet excitons in the organic pi-system, which are split at the organic-inorganic interface and quickly swept to the source-drain electrodes under a small bias. For comparison, we have also measured the 408 nm photocurrent response for OLm-capped $Cu_2Se_yS_{1-y}$ NCs, that is, without any CoTAPc sensitizers (see Supporting information **Figure S7**). Here, the responsivity is only on the order of 3 µA/W. We conclude that the dramatic increase in responsivity by almost 8 orders of magnitude observed after functionalizing $Cu_2Se_yS_{1-y}$ NCs with CoTAPc is due to the combined effect of better electronic coupling and the additional absorption of the organic pi-system. Thus, CoTAPc acts as an electronic cross-linker and optical sensitizer for the NCs.

## Conclusion

We have measured the structural, optical and electric properties of ternary $Cu_2Se_yS_{1-y}$ NC solids capped with oleylamine and the organic pi-system Cobalt β-Tetraaminophthalocyanine (CoTAPc), respectively, and compared it to the binary compound $Cu_2Se$. While the structural and optical response to ligand exchange and oxidation in air is rather similar for both materials, we have observed substantial differences in the charge carrier transport properties. Charge transport in $Cu_2Se_yS_{1-y}$ NC solids is dominated by electronic conduction, very sensitive to structural changes and largely unaffected by oxidation in air. Exchanging the surface ligand oleylamine with the organic π-system not only drastically increases electronic coupling in the $Cu_2Se_yS_{1-y}$ NC ensembles but also invokes an increase in the optical responsivity by eight orders of magnitude. Thus, ligand exchange with CoTAPc enables high conductivity and large responsivity in $Cu_2Se_yS_{1-y}$ NC films, which are much more robust against oxidation than their binary $Cu_2Se$ analogues.

## Acknowledgements

The authors acknowledge the DFG for support under Grants SCHE1905/3, DO1580/5 and SCHR700/25. This project has received funding from the European Research Council (ERC) under the European Union's Horizon 2020 research and innovation program (grant agreement No 802822). We also thank Mrs. Elke Nadler, Institute of Physical and Theoretical Chemistry, University of Tübingen, for performing SEM/STEM measurements using a Hitachi SU 8030 SEM which was funded by the DFG under contract INST 37/829-1 FUGG, partially.

## Corresponding Author:

*email: marcus.scheele@uni-tuebingen.de

## Author Contributions:

The manuscript was written through contributions of all authors. All authors have given approval to the final version of the manuscript.

## Supplementary Material:

(**S1**) XRD pattern of as-synthesized $Cu_2S_ySe_{1-y}$ NCs, (**S2**) AFM images and height profiles of NC films (**S3**) FT-IR spectra of $Cu_2Se$ and $Cu_2Se_yS_{1-y}$ NC films before and after ligand exchange with CoTAPc (**S4**) UV/VIS absorption spectrum of pure CoTAPc, (**S5**) I/V curves of CoTAPc-functionalized $Cu_{2-x}Se_yS_{1-y}$ NCs after varying exposure times to air, and (**S6**) Current–voltage characteristics of a $Cu_2Se$/OLm film after different oxidation/reduction treatments, and (**S7**) I/V curves of OLm-functionalized $Cu_{2-x}Se_yS_{1-y}$ NCs under photoexcitation by a 408 nm laser diode of varying optical power.

# Supporting Information

# Dye-Sensitized Ternary Copper Chalcogenide Nanocrystals: Optoelectronic Properties, Air Stability and Photosensitivity

*Sonam Maiti[1,2], Santanu Maiti[2], Ali Hossain Khan[3,4], Andreas Wolf[5], Dirk Dorfs[5], Iwan Moreels[3,4], Frank Schreiber[2,6] and Marcus Scheele[1,6]*

[1] Institute of Physical and Theoretical Chemistry, University of Tübingen, Auf der Morgenstelle 18, 72076 Tübingen, Germany.

[2] Institute of Applied Physics, University of Tübingen, Auf der Morgenstelle 10, 72076 Tübingen, Germany

[3] Istituto Italiano di Tecnologia, Via Morego 30, 16163 Genova, Italy

[4] Department of Chemistry, Ghent University, Krijgslaan 281-S3, 9000 Gent, Belgium

[5] Institute of Physical Chemistry and Electrochemistry, Leibniz Universität Hannover, Callinstr. 3A, 30167 Hannover, Germany

[6] Center for Light-Matter Interaction, Sensors & Analytics LISA+, University of Tübingen, Auf der Morgenstelle 15, 72076 Tübingen, Germany.

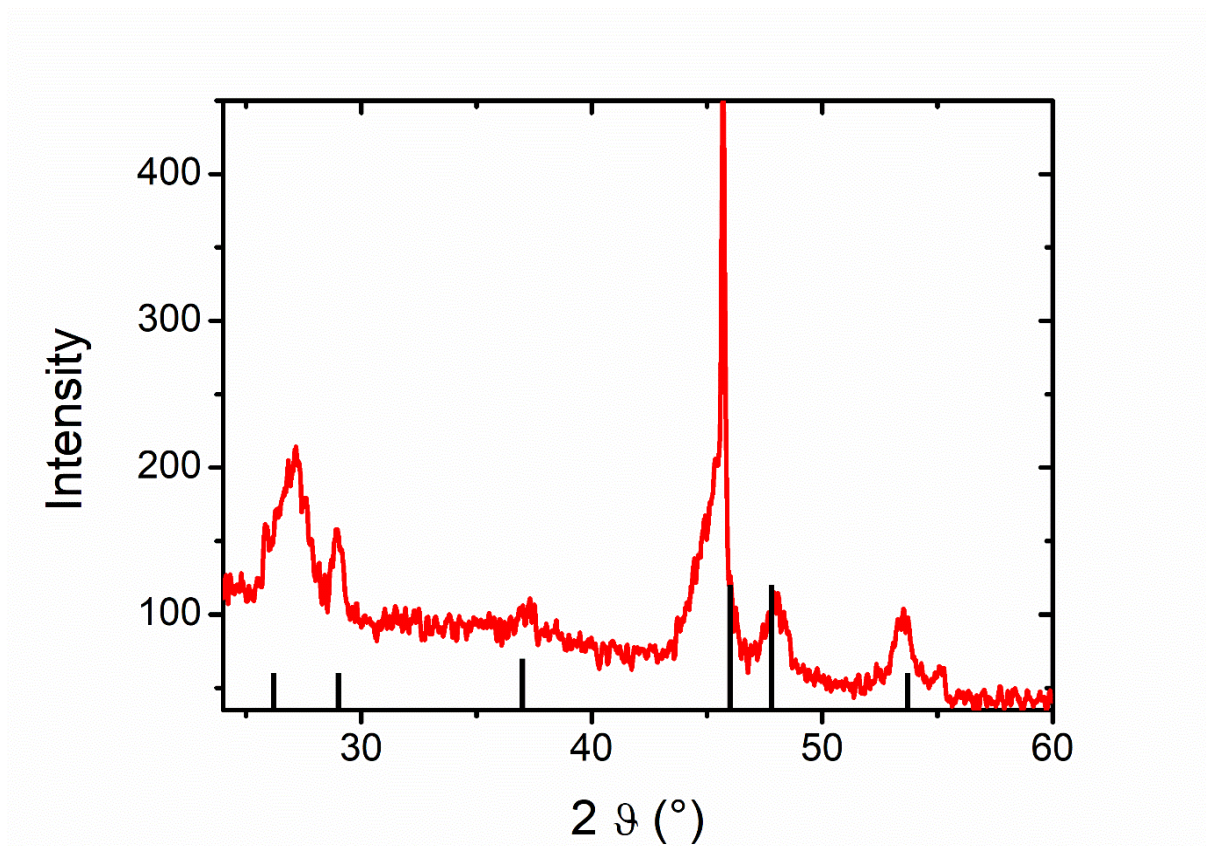

**Figure S1.** XRD pattern of as-synthesized $Cu_2S_ySe_{1-y}$ NCs. The reference structural data shows hexagonal $Cu_2S$ (PDF card #: 84-0209). The reflections at 29 °, 37 ° and 48 ° are absent in the cubic structure, and therefore characteristic for the hexagonal phase here.

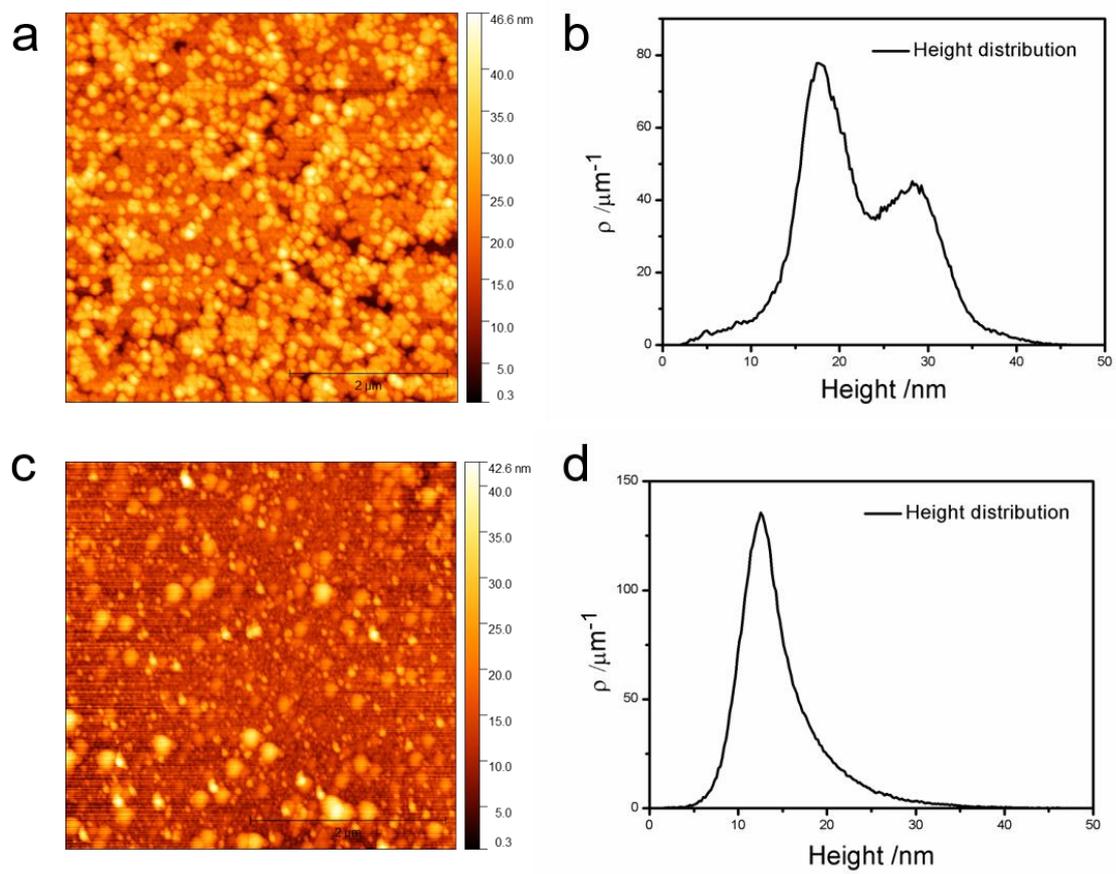

**Figure S2**. AFM images and extracted height profile of $Cu_2Se_yS_{1-y}$ (top) and $Cu_{2-x}Se$ (bottom).

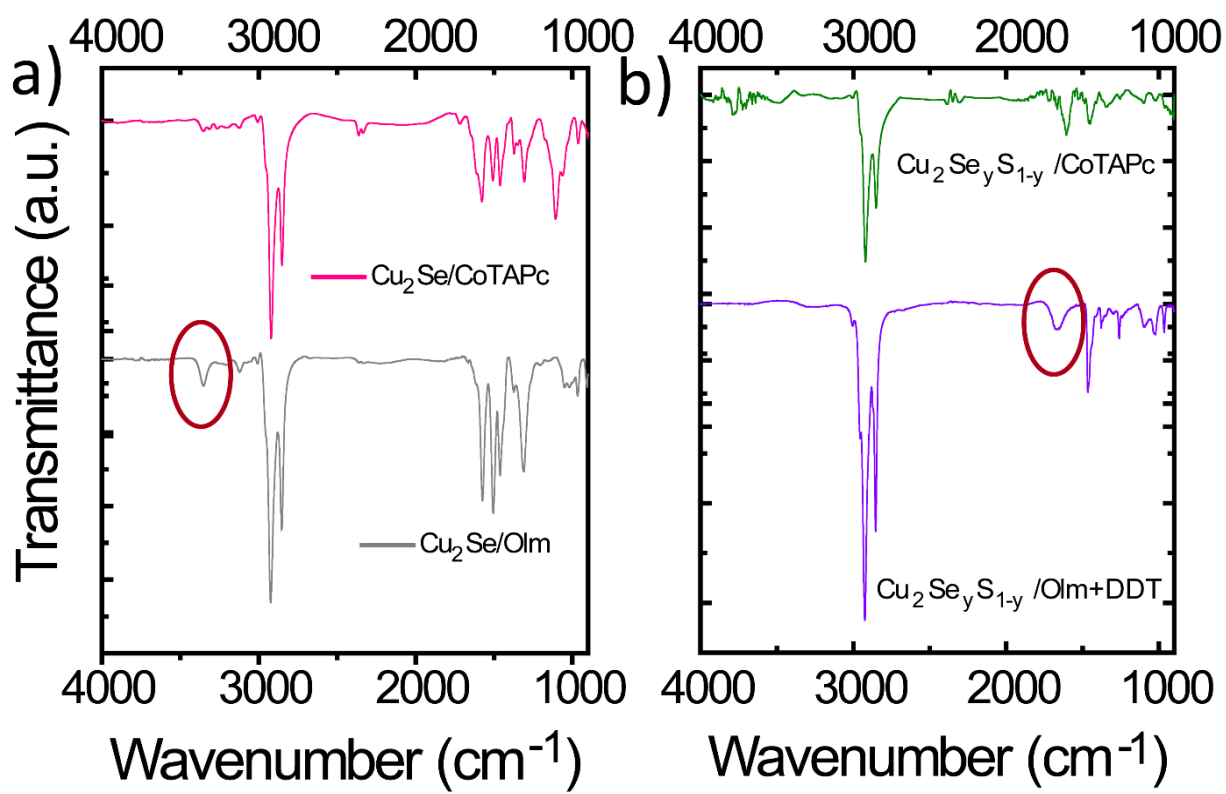

**Figure S3.** FT-IR spectra of **a)** Cu$_2$Se and **b)** Cu$_2$Se$_y$S$_{1-y}$ NC films, both before and after ligand exchange with CoTAPc. Characteristic bands for oleylamine, which disappear during ligand exchange have been encircled in red. Note: Since CoTAPc also contains -NH$_2$ and -CH vibrations, no quantitative description of the exchange efficiency can be made from this experiment.

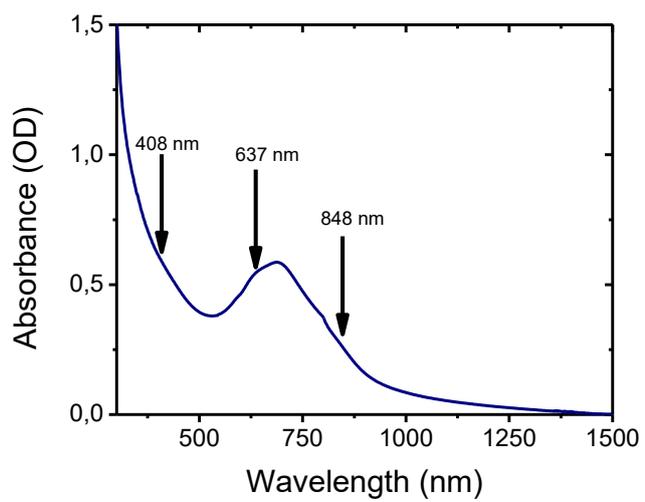

**Figure S4.** UV/VIS absorption spectrum of pure CoTAPc.

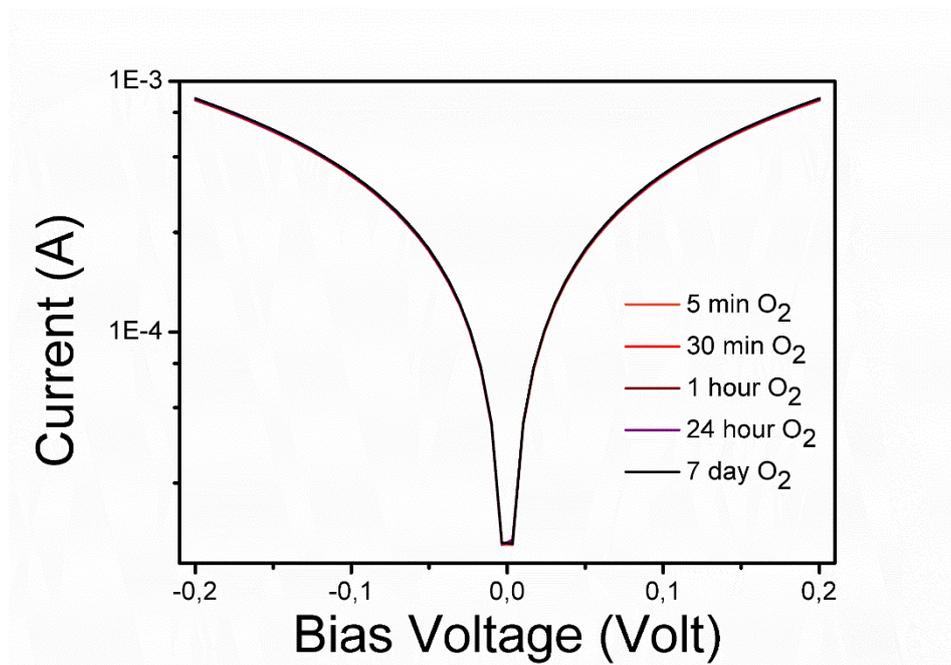

**Figure S5.** Current/Voltage characteristics of a $Cu_{2-x}Se_yS_{1-y}$/CoTAPc thin film at varying exposure times to air. The current was measured once the device was removed from the inert atmosphere of the glovebox and subsequently recorded from 5 minutes to 7 days.

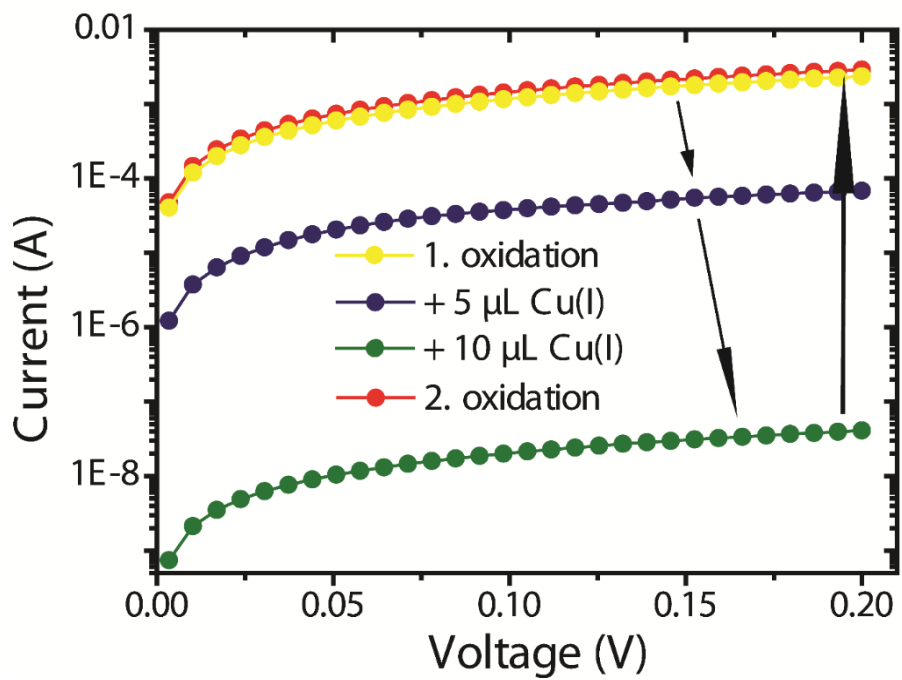

**Figure S6**. Current–voltage characteristics of a $Cu_2Se$/OLm film after a first oxidation in air (**yellow**), after treatment with 5 µL of a 0.04 mmol/L solution of $[Cu(CH_3CN)_4]PF_6$ (**blue**), after treatment with 10 µL of the same solution (**green**) and after 15 min of oxidation in air (**red**).

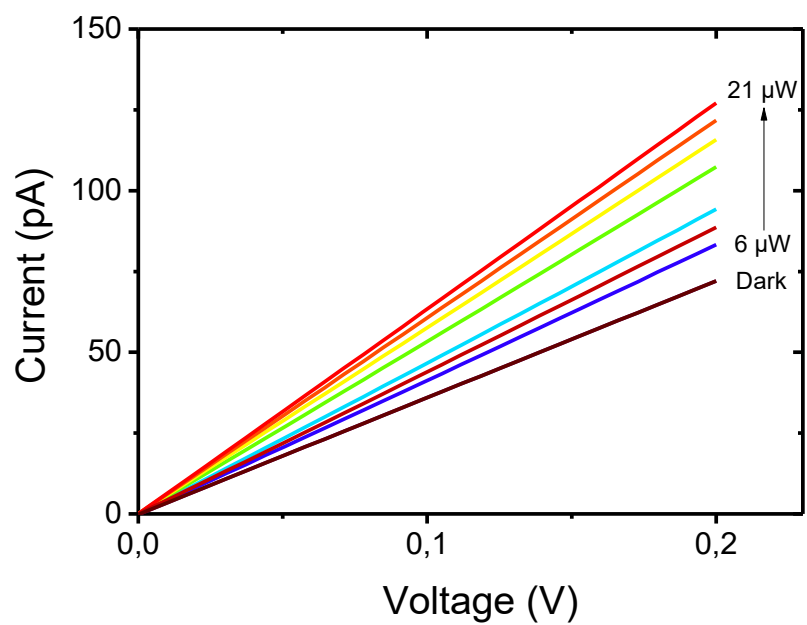

**Figure S7.** Current–voltage characteristics of a typical $Cu_{2-x}Se_yS_{1-y}$/OLm film at different absorbed optical power values provided by a 408 nm laser diode.